\documentclass[aps, prl, showpacs, twocolumn, 10pt, superscriptaddress, a4paper, floatfix]{revtex4-1}

\usepackage{graphicx}        
\usepackage{epstopdf}
\usepackage{CJK}
\usepackage{pifont}
\usepackage{bm}
\usepackage{amssymb}
\usepackage{amsmath}
\usepackage{siunitx}
\usepackage[usenames,dvipsnames]{xcolor}
\usepackage[linkcolor=violet,citecolor=blue,colorlinks=true,urlcolor=purple]{hyperref}

\newcommand{\new}[2][]{#2}

\newcommand{\Letter}{{Article}}

\newcommand{\ntwo}{N$_2$}
\newcommand{\otwo}{O$_2$}

\renewcommand{\vec}{\textbf}

\begin{document}
\begin{CJK*}{UTF8}{gbsn}
\title{Exploring photoelectron angular distributions emitted from molecular dimers by two delayed intense laser pulses}

\author{V\'{a}clav\,Hanus}
\email[]{vaclav.hanus@tuwien.ac.at}
\affiliation{Photonics Institute, Technische Universit\"at Wien, 1040 Vienna, Austria, EU}
\affiliation{Wigner Research Centre for Physics, Institute for Solid State Physics and Optics, Budapest, Hungary, EU}

\author{Sarayoo\,Kangaparambil}
\affiliation{Photonics Institute, Technische Universit\"at Wien, 1040 Vienna, Austria, EU}

\author{Seyedreza\,Larimian}
\affiliation{Photonics Institute, Technische Universit\"at Wien, 1040 Vienna, Austria, EU}

\author{Martin\,Dorner-Kirchner}
\affiliation{Photonics Institute, Technische Universit\"at Wien, 1040 Vienna, Austria, EU}

\author{Xinhua\,Xie (谢新华)}
\affiliation{Photonics Institute, Technische Universit\"at Wien, 1040 Vienna, Austria, EU}
\affiliation{SwissFEL, Paul Scherrer Institute, 5232 Villigen PSI, Switzerland}

\author{Andrius\,Baltu\v{s}ka}
\affiliation{Photonics Institute, Technische Universit\"at Wien, 1040 Vienna, Austria, EU}

\author{Markus\,Kitzler-Zeiler}
\email[]{markus.kitzler-zeiler@tuwien.ac.at}
\affiliation{Photonics Institute, Technische Universit\"at Wien, 1040 Vienna, Austria, EU}

\begin{abstract}
We describe the results of experiments and simulations performed with the aim of extending   photoelectron spectroscopy with intense laser pulses to the case of molecular compounds.  
Dimer frame photoelectron angular distributions generated by double ionization of N$_2$-N$_2$ and N$_2$-O$_2$ van der Waals dimers with ultrashort, intense laser pulses  are measured using four-body coincidence imaging with a reaction microscope. 
To study the influence of the first-generated molecular ion on the ionization behavior of the remaining neutral molecule we employ a two-pulse sequence comprising of a linearly polarized and a delayed elliptically polarized laser pulse that allows distinguishing the two ionization steps.
By analysis of the obtained  electron momentum distributions we show that scattering of the photoelectron on the neighbouring molecular potential leads to a deformation and rotation of the photoelectron angular distribution as compared to that measured for an isolated molecule. 
Based on this result we demonstrate that the electron momentum space in the dimer case can be separated, allowing to extract information about the ionization pathway from the photoelectron angular distributions. 
Our work, when implemented with variable pulse delay,  opens up the possibility of   investigating light-induced electronic dynamics in molecular dimers using angularly resolved photoelectron spectroscopy with intense laser pulses.
%
\end{abstract}





\maketitle
\end{CJK*}



\section{Introduction}

A powerful method for tracing molecular dynamics induced by a pump pulse is to observe the evolution of the angular distributions of photoelectrons emitted by single- or few-photon ionization during a delayed probe pulse \cite{Stolow2004, Hertel2006}. If combined with coincidence imaging of the photoion momenta, this technique allows for the measurement of molecular frame photoelectron angular distributions (MFPADs) of a single molecule, which can provide unique insight into the intra-molecular dynamics with femtosecond resolution \cite{Gessner2006}. 
\new[R1-1 ]{This method can also be implemented with an intense, elliptically polarized laser pulse as the ionizing probe \cite{Akagi2009, Staudte2009, Odenweller2011, Spanner2012, Wu2012, Hanus2020}.
In that case, the rotating electric field vector of the probe pulse can be exploited for mapping laser-sub-cycle time to electron momentum, a concept known as angular streaking \cite{Maharjan2005, Eckle2008a, Schoffler2016, Hanus2019, Kangaparambil2020, Hanus2020}. 
%
When distortions of the such obtained MFPADs due to the parent ion's Coulomb potential \cite{Staudte2009, Spanner2012}, the field-driven electronic dynamics \cite{Odenweller2011}, or the nuclear dynamics \cite{Hanus2019, Hanus2020} that take place concomitantly with the photoelectron emission are properly accounted for, molecular dynamics can be extracted from them with sub-femtosecond resolution.}

The next frontier in ultrafast intense laser science is to extend existing methods that are now routinely applied to isolated molecules, to the study of dynamics in more complicated systems such as molecules in a compound.  This effort is motived by the fact that most molecular processes in nature or technical applications do not take place within a single molecule isolated in vacuum but between different molecules. A famous example for such an inter-molecular process is proton coupled electron transfer \cite{Weinberg2012}, which is of key importance in biology, chemistry, and technology. 
A widely used approach to study photoinduced inter-molecular processes in femtosecond pump-probe experiments  is to use small molecular van der Waals clusters and dimers as model systems \cite{Stolow2004, Hertel2006}. 
In recent years, atomic and molecular dimers have also found increasing attention in experiments that study their ionization and fragmentation dynamics when driven by a strong laser fields \cite{Manschwetus2010, Ulrich2010, Wu2011, Ulrich2011, Wu2011b, Wu2012d, Hoshina2012, Wu2012b, VonVeltheim2013, Wu2014a, Amada2015, Ding2017, Cheng2017, Wang2020}. 
It was demonstrated that the analysis of the momenta of photoions or photoelectrons emitted during ionization-fragmentation of dimers can provide detailed insight into, e.g., the ionization dynamics 
\cite{Wu2013b}, 
the structural deformation induced by the strong-field interaction 
\cite{Amada2015}, 
the influence of a nearby charge on the dissociation behaviour of molecules
\cite{Ding2017}, 
or into laser-sub-cycle electron transfer processes 
\cite{Wang2020}. 
A common finding of these studies is that during multiple ionization the nearby charge from the neighbouring ion can have a decisive impact on the further ionization and/or fragmentation dynamics of the partner molecule. 


In this \Letter{} we describe experiments and simulations performed with the aim of investigating to what extent photoelectron angular distributions (PADs) measured with a strong, elliptically polarized laser field can be used to extract structural and dynamical information from molecules bound in a hetero-dimer complex. To this end we studied with a reaction microscope \cite{Ullrich1997, Doerner2000, Hanus2019, Hanus2020, Kangaparambil2020} 
the PADs from sequential double ionization of \ntwo{}-\ntwo{} and \ntwo{}-\otwo{} dimers during two delayed intense laser pulses 
in coincidence with the photoions ejected during subsequent fragmentation of the dimers. 
Our main concern was to understand which information about the laser-molecule interaction is contained in the measured PADs and how this information can be extracted. 
The double-pulse approach in combination with the coincidence measurement permitted us to reconstruct PADs in the dimer frame of reference (referred to as DFPADs) specifically for the second ionization step. 
This allowed us to identify the influence of a nearby molecular ion on the ionization and fragmentation behaviour of a neutral molecule in measured DFPADs.

From our measurements we find that the nearby charge of the first-created molecular ion mainly leads to a rotation of the measured photoelectron angular distribution as compared to the isolated molecule case, whereas its overall shape is to a large degree preserved. 
Thus, one of the results of this study is that DFPADs can, with adaptations, be read in a similar way as MFPADs. 
Furthermore, by applying methods based on angular streaking developed previously \cite{Hanus2019, Kangaparambil2020, Hanus2020} and with the support from semi-classical trajectory simulations we are able to show that the dominant process underlying the rotation is scattering of the second-emitted electron on the neighbouring molecular ion as it is driven away by the strong laser field.
This finding adds to the emerging evidence for the key importance of laser-sub-cycle electron scattering dynamics in strong-field driven dimers \cite{Wang2020} and clusters \cite{Treiber2020}. 





\begin{figure}[tb]
  \centering
  \includegraphics[width=1\columnwidth]{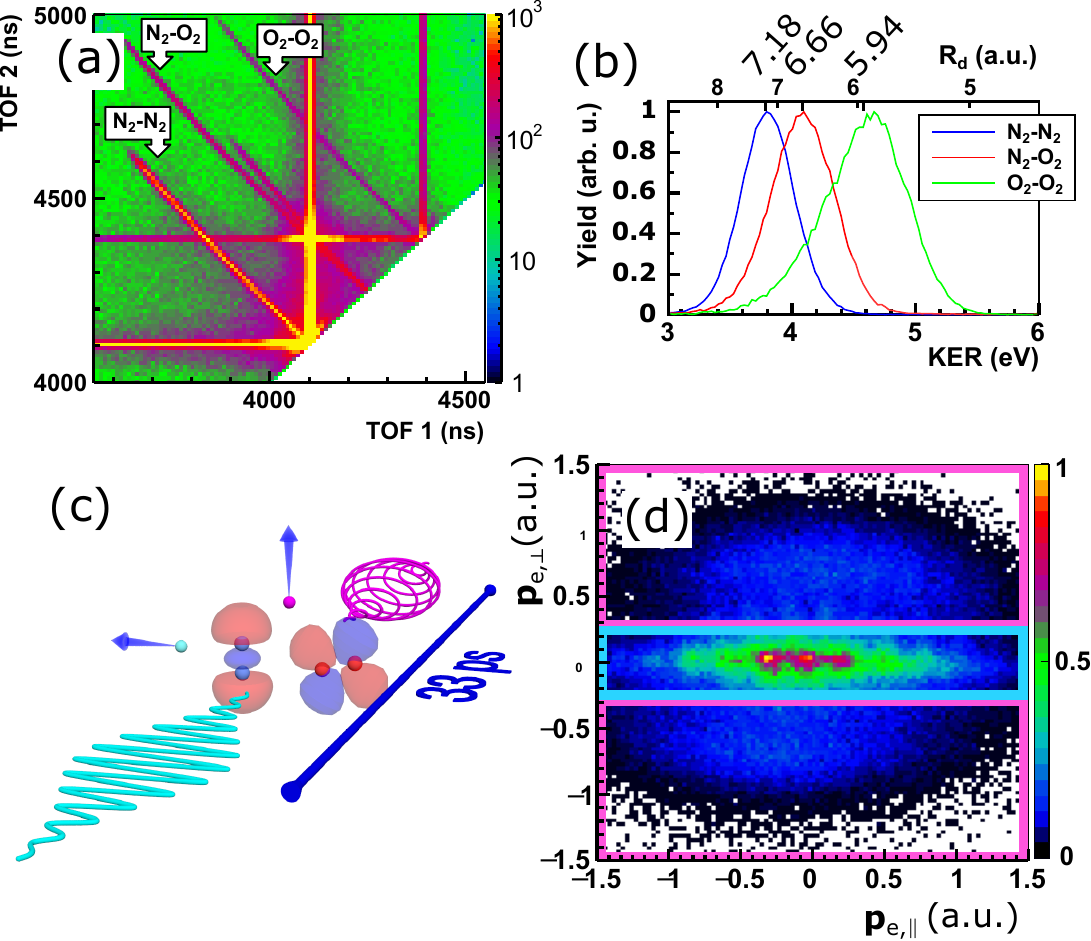}
  \caption{
(a) Measured photoion-photoion coincidence (PIPICO) distributions \new[R1-3, R2-5 ]{obtained by application of both the pump and probe pulses}. Indicated are the three dimer species that are generated and detected in our experiment.
(b) Kinetic energy distributions (KER) that result from Coulomb explosions initiated by laser double ionization of the three dimer species shown in (a) \new[R1-3, R2-5 ]{when both the pump and probe laser pulses were applied}.
(c) Schematic representation of the experiment. A linearly polarized and an elliptically polarized pulse delayed by 33\,ps each remove one electron \new[R1-4 ]{(cyan and pink electric field-vector traces)} from one of the two molecules in the dimer. Depicted is the example of a dimer consisting of an N$_2$ molecule with $\sigma$-geometry of the highest occupied molecular orbital (HOMO) and an O$_2$ molecule with a $\pi$-geometry of the HOMO.
(d) Measured electron momentum distribution in the laser polarization plane for N$_2$ after interaction with the double-pulse sequence. The distributions for the dimers look similar.
The narrow elongated structure in the cyan box corresponds to electrons emitted during the linearly polarized pulse, for which electron emission happens dominantly along the laser polarization axis [cf. cyan sphere in panel (c)]. The arc segment-shaped parts in the pink boxes are due to electron emission in the elliptically polarized delayed pulse [cf. pink sphere in panel (c)].}
  \label{fig1}
\end{figure}

\section{Experiments}

The reaction microscope used in our experiments to measure in coincidence the momenta of two molecular ions and two electrons emerging from the interaction of a sequence of two intense laser pulses with a cold jet of molecules generated by ultrasonic expansion of N$_2$ and O$_2$ gas is described elsewhere \cite{Hanus2019, Hanus2020, Kangaparambil2020}. 
The laser pulses used in the experiments had both a central wavelength of $790$\,nm and a duration of about \SI{25}{\fs}. 
\new[R1-2 ]{
Since the gas is strongly cooled during the ultrasonic expansion,  molecular dimers, as the pre-cursors of a quantum fluid, may be formed during the ultrasonic expansion \cite{Ullrich1997, Doerner2000}. 
When these dimers interact with the intense laser pulses, each molecule in the dimer might become singly ionized, leading to Coulomb explosion of the doubly ionized dimer complex. 
The signatures of these Coulomb explosions are shown in the photoion-photoion coincidence (PIPICO) histogram in Fig.~\ref{fig1}(a) as the sharp lines. These lines  reflect the momentum conservation between the two exploding fragment ions and therefore are clear signatures existence of a certain  molecular species in the cold gas jet. The histogram in Fig.~\ref{fig1}(a) shows that in our experiment N$_2$-N$_2$, O$_2$-O$_2$ and N$_2$-O$_2$ dimers were formed during the ultrasonic expansion and Coulomb exploded during the interaction with the strong laser pulses. 
}
The laser interaction took place in an ultra-high vacuum chamber (base pressure 0.9$\times$10$^{-10}$\,mbar).  Ions and electrons emerging from the interaction volume were guided by weak electric (19\,V/cm) and magnetic fields (12\,G) to two separate position-sensitive detectors. From the time-of-flight and impact position of each detected particle its  momentum right after the laser interaction was calculated.

\begin{figure*}[t]
  \centering
  \includegraphics[width=2\columnwidth]{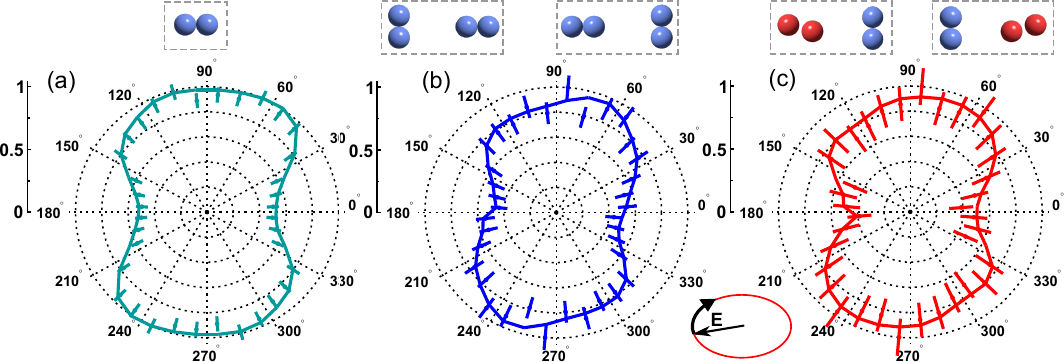}
  \caption{Measured molecular frame and dimer frame photoelectron angular distributions (MFPADs respectively DFPADs) \new[R1-5, R2-1c ]{with statistical error bars for the (a) N$_2$ molecule, (b) N$_2$-N$_2$ dimer and (c) N$_2$-O$_2$ dimer. The lines are to guide the eye and represent fits to the measured data. }
Depicted are the angular distributions of the second electron released during the elliptically polarized pulse. For the DFPAD in (c) the electrons for both orientations of the N$_2$-O$_2$ dimer were superposed.
The molecular respectively dimer axes, retrieved from the ions' momenta, are aligned horizontally, which coincides with the major axis of the polarization ellipse, see the cartoon between (b) and (c), which also depicts the clockwise rotation of the electric field vector. }
  \label{fig2}
\end{figure*}

We optimized the expansion conditions for a high production rate of N$_2$-O$_2$ dimers in the jet, since we were interested in the influence of the molecular species on the ionization and fragmentation dynamics. 
Hence, O$_2$ was set to have lower abundance in the gas jet as compared to N$_2$. This led to low yield of O$_2$-O$_2$ dimers and, thus, to low statistic in their DFPADs. Therefore, in the following, we will not discuss angularly resolved data for  O$_2$-O$_2$ dimers.

During Coulomb explosion of a dimer into two charged molecular ions with masses $m_{1,2}$, the electrostatic potential energy is released as kinetic energy of the two ions with momenta $\vec{p}_i$. The  kinetic energy released (KER) during this explosion, $E_{\mathrm{KER}}=\sum_i \vec{p}_i^2/(2 m_i)$, is a precise measure of the intermolecular distance, $R_d = 1/E_{\mathrm{KER}}$, of the dimer \cite{Vager1989}.
The measurement revealed, cf.\,Fig.~\ref{fig1}(b), that $E_{\mathrm{KER}}$ is the largest for the O$_2$-O$_2$ dimer, hence it exhibits the smallest intermolecular distance of 5.94\,a.u. The smallest energy release was measured for the N$_2$-N$_2$ dimer, corresponding to a van~der~Waals bond length of 7.18\,a.u. 
An intermediate distance of $R_d = 6.66$\,a.u. was measured for the heteronuclear dimer  N$_2$-O$_2$.

\new[R1-3, R2-5 ]{The KER distributions in Fig.~\ref{fig1}(b) were obtained when both laser pulses of the double-pulse sequence were used for Coulomb exploding the dimers. As we will describe in detail below, the combined use of a linearly polarized pump pulse and a preceding circularly polarized probe pulse, delayed to the pump by 33\,ps, cf. sketch in Fig.~\ref{fig1}(c), allowed us to distinguish the two possible ionization sequences that can initiate the fragmentations leading to the KER distributions in Fig.~\ref{fig1}(b). These two possibilities are: (i) both ionization events take place within the duration of one of the two applied pulses (i.e., within a few tens of femtoseconds), (ii) one ionization event takes place during the pump pulse, the second one during the probe pulse (delay about 33\,ps). Exploiting this opportunities we found that the measured KER distributions corresponding to the two possibilities agree within experimental errors [not shown in Fig.\,\ref{fig1}(b)]. This indicates that the dimers' bond stretch dynamics is negligible during the 33\,ps delay. Thus, the most likely scenario is that the potential energy curves of the singly charged dimers are binding and of very similar shapes as those of the neutral dimers.
}

Because in a two-body Coulomb explosion the two particles are ejected back-to-back, the measured ions' momenta $\vec{p}_1=-\vec{p}_2$ directly reflect the orientation of the dimer axis in the laser focus at the time of interaction, provided the explosion happens much faster than the rotational dynamics that might be induced during the laser interaction \cite{Tong2005b}. As for the large size and mass of the dimers this is well fulfilled, the dimer axis and therewith DFPADs can be reconstructed from the measured electron and ion momenta.

\section{Double-pulse approach to distinguish the two ionization steps}

Our goal was to measure the DFPAD generated by an elliptically polarized laser pulse when one of the two molecules in the dimer was ionized by a preceding pulse. 
For such measurement it is necessary to discriminate the first emitted from the second emitted electron in the detected electrons.
This discrimination cannot be simply made based on the flight times of the two electrons detected in coincidence, as the first detected could be the second emitted electron and vice versa.
Therefore, we employed a technique that exploits the different shape of the electron momentum distributions produced by linearly and elliptically polarized laser pulses \cite{Song2016}: Electrons released in linearly polarized light exhibit small momenta perpendicular to the light polarization direction and therefore cover regions in an electron momentum distribution  that are dominantly aligned along the polarization axis. In contrast, electrons released in elliptically polarized light are angularly streaked and cover momentum regions with large perpendicular momentum that resemble circle segments \cite{Schoffler2016, Hanus2019, Hanus2020}. Thus, it becomes possible to minimize the momentum space overlap between two electrons released in a two-pulse sequence consisting of a linearly polarized pump laser pulse and a delayed, elliptically polarized probe pulse, when during each pulse only one electron is emitted, cf. sketch in Fig.~\ref{fig1}(c). In our experiments the helicity of the elliptically polarized probe pulse was chosen such that the electric field vector rotates clockwise in the electron momentum plane. 
The delay between the two pulses was set constant to \SI{33.3}{\pico\s} in order to ensure that the pulses do not overlap and that the prompt alignment of dimers or possible revivals of the alignment are avoided \citep{Xie2014a, Kumarappan2007, Wu2011b}.

The pulse peak intensities of the two pulses were chosen as \SI{4e14}{\W\per\cm\squared} for both pulses. The intensities were calibrated \textit{in situ} using the proton distribution from the dissociation of hydrogen, H$_2$, from the background gas in the ultra-high vacuum chamber \cite{Alnaser2004}.
For these intensities, chosen because they are typical for strong-field experiments that investigate laser-ionization, an optimization of the ellipticity that produced minimal  overlap of the two electron momentum distributions from each pulse yielded $E_\perp/E_\parallel=0.6$. Here, $E_\perp$ and $E_\parallel$ are the laser peak field strengths perpendicular and parallel to the main axis of the polarization ellipse of the elliptically polarized pulse, which was aligned with the polarization direction of the linearly polarized laser pulse. 
Throughout the \Letter{} this axis is assumed along the horizontal direction. 
The electron momentum distribution measured with these pulse sequences is shown in Fig.~\ref{fig1}(d) exemplary for N$_2$. The distributions for the dimers look similar. 

Provided that all ions and electrons are detected in coincidence, as was done in our experiment, 
the double-pulse approach enables the selection of dimer fragmentation events 
when the first electron was \emph{exclusively} released by the pump pulse and the second electron \emph{exclusively} by the probe pulse. 
To reconstruct the MFPADs respectively DFPADs corresponding to the second ionization step, 
we separated the measured electron momentum distributions in the laser polarization plane into three regions using $\vec{p}_{e,\perp}$, the electrons' momentum perpendicular to the polarization direction of the linear pulse, as the decisive parameters, see colored boxes in Fig.~\ref{fig1}(d). 
Electrons with $|\vec{p}_{e,\perp}|\leq 0.3$\,a.u. [cyan box in Fig.~\ref{fig1}(d)] were considered to be emitted during the linearly polarized first pulse. Electrons with $|\vec{p}_{e,\perp}|> 0.3$\,a.u. [pink boxes in Fig.~\ref{fig1}(d)] were considered to be emitted during the elliptically polarized second pulse. 
We only considered those coincidence events, where one of the two electrons  was in the cyan-bordered momentum region and the other electron in one of the two pink-bordered regions. All other coincidence events were discarded. 
\new[R2-1c ]{This separation works perfectly only for negligible overlap of the electron momentum distributions due to the linear and elliptical pulse, respectively. In practice, a small overlap of the wings of two momentum distributions is unavoidable. 
Three main types of systematic errors resulting from this overlap can be distinguished: (i) The electron emitted during the elliptical pulse is emitted with $|\vec{p}_{e,\perp}|\leq 0.3$\,a.u., (ii) the electron emitted during the linear pulse is emitted with $|\vec{p}_{e,\perp}|> 0.3$\,a.u., (iii) both electrons are emitted during the elliptical pulse, one with $|\vec{p}_{e,\perp}|> 0.3$\,a.u., the other one with $|\vec{p}_{e,\perp}|\leq 0.3$\,a.u. By fitting the three main lobes in the electron momentum distributions [that all have a similar shape as the one in  Fig.\,\ref{fig1}(d)] with three Gaussians and exploiting coincidence arguments, we found that only case (iii) can result in noteworthy errors. However, even for this case this is only problematic, when the first emitted electron reaches $|\vec{p}_{e,\perp}|> 0.3$\,a.u. and the second emitted only $|\vec{p}_{e,\perp}|\leq 0.3$\,a.u. But as the first emission happens most likely during the onset and rising slope of the pulse, while the second emission is expected to happen around its peak, even this error has a small probability. In sum, we found that the double-pulse method with well-chosen relative peak intensities, as it is the case in our experiment, is relatively insensitive to the unavoidable overlap of the wings of the electron momentum distributions.
}

\new[R2-1a ]{
From the coincidence events that remained after demanding that the first  electron was emitted during the linear pulse (with $|\vec{p}_{e,\perp}|\leq 0.3$\,a.u.) and the second one during the elliptical pulse (with $|\vec{p}_{e,\perp}|> 0.3$\,a.u.), we furthermore discarded those events where the ejection direction of the ionic fragments was perpendicular to the major axis of the polarization ellipse: We only considered those coincidence events where the internuclear axis in the case of molecules, respectively the intermolecular axis in the case of dimers, were oriented within an angle of $\pm45^\circ$ to the major axis of the polarization ellipse. The orientation of the molecules/dimers was calculated from the fragment ions' momentum vectors $\vec{p}_1=-\vec{p}_2$.
The reason for this selection was motivated by our recent work Ref.~\cite{Wang2020} in which we found that the ionization behaviour of a dimer is strongly influenced by laser-driven electron scattering and transfer processes between the two entities of the dimer. 
To investigate the influence of such electron scattering processes with the double-pulse method, the dimer axis should be roughly aligned along the dominant ionization direction where the field strength is largest, i.e., the major axis of the polarization ellipse. 
If the dimers' axes were aligned perpendicular to this direction (i.e., along the minor axis), scattering of the ionizing electron on the opposing molecular entity is expected to be of less importance. 
Again, all coincidence events where the dimers/molecules were not aligned within $\pm45^\circ$ to the major axis were discarded.

For the remaining coincidence events, one of the two detected ions (either N$^+$, N$_2^+$, or O$_2^+$, depending on the molecule/dimer considered) was taken as the reference fragment ion and its momentum vector was rotated, together with the electron momentum vectors, into the positive horizontal axis. Then, we calculated for the electrons in the pink boxes in Fig.~\ref{fig1}(d), that are attributed to emissions during the elliptically polarized probe laser pulse, the angle between the reference ion and the electron momentum vector, denoted by $\alpha$ in the following. 
Plotting the electron yield as a function of $\alpha$ yielded, finally, the MFPADs/DFPADs shown in Fig.~\ref{fig2}. 
}

\begin{figure*}[t]
  \centering
  \includegraphics[width=2\columnwidth]{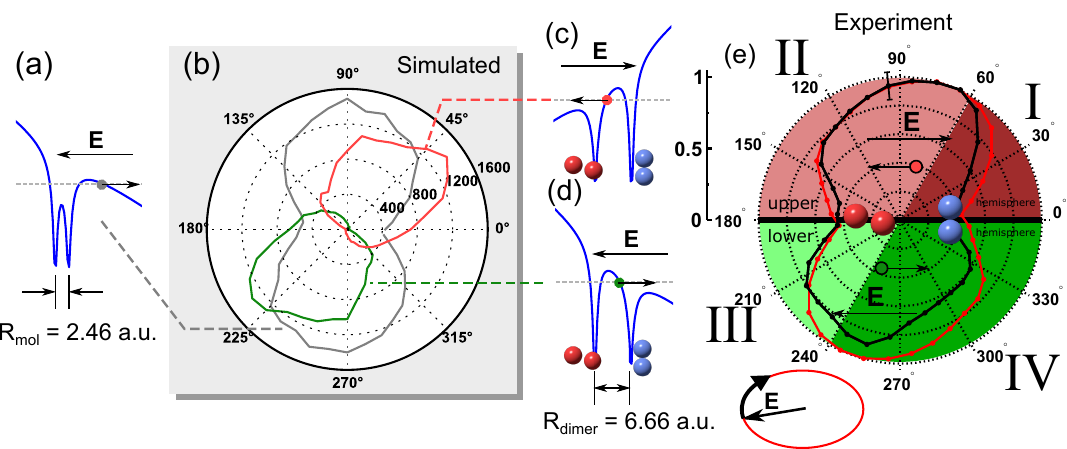}
  \caption{(a)-(d) Results of simulations detailed in the text and explanatory sketches. The grey line in (b) shows the simulated angular distribution of photoelectrons emitted from a molecular-type potential depicted in (a). The assumed internuclear distance of 2.46\,a.u. corresponds to that of neutral N$_2$. The red line in (b) shows the photoelectron angular distribution when the electrons tunnel from the up-field potential well towards the left. This situation, depicted in (c) with a red circle indicating the tunneling electron, occurs when the laser electric field vector points to right (as indicated by an arrow). 
The green line in (b) shows the angular distribution for the opposite case, sketched in (d), when the electric field vector points to the left and the electron (green circle) is ejected to the right. 
(e) Measured angular distribution of the photoelectron emitted from the N$_2$-O$_2$ dimer during the elliptically polarized probe pulse. 
The black line shows the data when the dimer is oriented such that N$_2$ is on the right side.  \new[R2-1c ]{For clarity, only one representative statistical error bar is shown (around 90$^\circ$).} 
The red line is added for reference and shows the angular distribution of the non-oriented dimer, duplicated from Fig.\,\ref{fig2}(c). See text for further details. 
}
  \label{fig3}
\end{figure*}

As for the N$_2$ molecule and the N$_2$-N$_2$ dimer the two ions are indistinguishable, we mirrored the MFPADs/DFPADs about the vertical axis. The same was done for the N$_2$-O$_2$ dimer in Fig.~\ref{fig2}(c), yielding a DFPAD where both orientations (N$_2$ ejected to the right, and N$_2$ ejected to the left) are superposed. A DFPAD for the N$_2$-O$_2$ dimer, where the electron and ion momenta were rotated together such that the momentum vector of N$_2$ exclusively points to the right, is shown in Fig.~\ref{fig3}(e) and will be discussed in detail below.

\section{Analysis of measured dimer-frame photoelectron angular distributions (DFPADs)}

To understand the information contained in DFPADs, 
we first compare them to the more familiar case of an MFPAD measured for a Coulomb exploding doubly ionized molecule. 
Fig.~\ref{fig2}(a) shows the MFPAD for the N$_2$ molecule measured with the double-pulse sequence, where the first ionization event takes place during the linearly polarized pump pulse and the second one during the elliptically polarized probe pulse. 
Figs.~\ref{fig2}(b) and \ref{fig2}(c) show the DFPADs for the N$_2$-N$_2$ and N$_2$-O$_2$ dimers measured with exactly the same pulse sequence.  The MFPAD features a symmetric shape about the symmetry axis  along $\alpha=85^\circ$, with $\alpha$ the angular coordinate. The small ca. 5$^\circ$-rotation from $\alpha=90^\circ$ that would be expected in angular streaking for an elliptically polarized pulse with its major axis of the polarization ellipse along $\alpha=0^\circ$,  may be attributed to the influence of the Coulomb field on the emitted electron \cite{Spanner2012}.
In contrast to the symmetric MFPAD, the DFPADs feature a non-symmetric shape about the $\alpha=85^\circ$ axis: For the upper hemisphere, $\alpha=[0^\circ, 180^\circ]$, the DFPADs contain a larger number  of events in the range $\alpha<90^\circ$. This asymmetry is even more pronounced in the DFPAD measured for N$_2$-O$_2$ shown in Fig.~\ref{fig2}(c). 

To explain the reason for the different shapes of the MFPAD and the DFPADs depicted in Fig.~\ref{fig2} we  turn to discussing the differences in the binding potentials from which the second electron is detached. 
The photoelectron distributions plotted in Fig.~\ref{fig2} reflect the angular dependence of the second ionization step. In a simplified picture we can assume that the second electron,  at the instant of its emission, is bound in a potential well that is defined by two positive charge centres. In the case of a molecule these centres are close to each other and there is no considerable barrier between them, see the sketch in Fig.\,\ref{fig3}(a). When the electron tunnels through the field-suppressed barrier it directly reaches the continuum. If the two centres are further apart, as it is the case for a dimer, two situations can be distinguished: The electron can be either released from the  down-field or the up-field well. An electron released from the down-field can tunnel directly to the continuum similar to the molecular case sketched in Fig.~\ref{fig3}(a). In contrast, when an electron tunnels from the up-field well, as sketched in Figs.~\ref{fig3}(c) and \ref{fig3}(d), there is a possibility that the electron, as it is driven away by the laser electric field, scatters on the down-field charge centre.

In the following we will show by an in-depth analysis of the measured DFPADs, supported by results of semiclassical trajectory Monte Carlo simulations,   that the asymmetry in the measured DFPADs in Figs.\,\ref{fig2}(b) and \ref{fig2}(c) is indeed caused by scattering of electrons released from the up-field well on the opposing charge center.
\new[R2-1b, R2-2 ]{To guide our analysis, we simulate the momentum distribution of the electron that is emitted during the elliptically polarized pulse from a model binding potential by tunnel ionization and subsequently is driven by the combined forces due to the laser electric field and a binding potential situated at some distance $R$ from the electron's parent binding potential.
The neighbouring binding potential mimics the positively charged ion created by ionization during the preceding linearly polarized pump pulse. The electron released during the linear pulse is not simulated and it is assumed that this electron has already left the interaction volume. This is perfectly justified given the delay of 33\,ps between the two pulses. 
Thus, the potential felt by the ionizing and laser-driven electron has a Coulomb double-well shape. To avoid numerical problems due to the Coulomb singularity we used soft-core potentials of the form
\begin{equation}
V(\vec{r}) = - \sum_{i=1}^2 \frac{1}{\sqrt{(\vec{r}-\vec{R}_i)^2 + A}},
\end{equation}
where is $A$ the Coulomb softening parameter that was set to 0.1 for all calculations, and $\vec{r}$ is the position of the simulated electron. At position $\vec{R}_{1}$ the singly charged neighbouring ion is situated, at $\vec{R}_{2}$ is the parent ion of the electron. 
The separation $R=|\vec{R}_{1}-\vec{R}_{2}|$ of the two ions was chosen as $R=2.46$\,a.u. for the N$_2$ molecule and $R=6.66$\,a.u. for the N$_2$-O$_2$ dimer. 
At every instant $t_i$ of the discrete temporal mesh the starting position of the electron, $r_0$, i.e., the tunnel exit sketched in Figs.\,\ref{fig3}(a,c,d), was calculated by numerically solving the equation
%
\begin{equation}
V(\vec{r}_0) + \vec{r}_0\cdot\vec{E}(t_i) = -|I_p| 
\end{equation}
using the secant method. Here, $\vec{E}(t)$ is the vectorial laser field and $I_p$ is the assumed ionization potential. At each instant of the temporal grid a trajectory was launched at the position $\vec{r}_0(t_i)$ with a probability given by the molecular ADK ionization theory \citep{Tong2002}.
As the sole purpose of the simulations was to obtain a qualitative understanding of the ionization behaviour and the subsequent laser-driven free electron dynamics rather than obtaining quantitative agreement with the measured data, we used, for the sake of simplicity, the parameters of the helium atom in the ADK formula, where the $I_p$ was adapted to that of the molecules.
The trajectories were propagated in three dimensions for an elliptically polarized laser pulse with the same parameters as in the experiments using the Runge-Kutta integration scheme. The helicity of the elliptically polarized laser pulse was, as in the experiments, assumed clockwise. 
From the final momentum value $\dot{\vec{r}}_\infty$ 
of each trajectory at time $t\rightarrow \infty$ (long after the pulse) the electron momentum distribution was obtained by binning all launched trajectories.
}

\begin{figure}[tb]
  \centering
  \includegraphics[width=1\columnwidth]{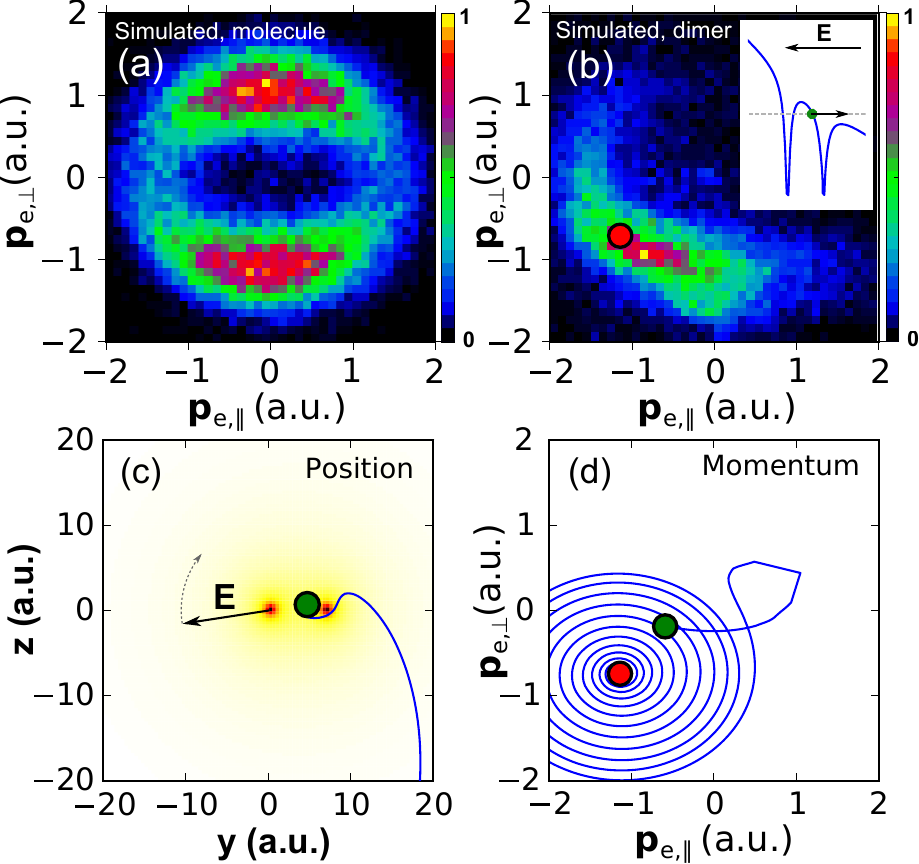}
  \caption{Simulated electron momentum distributions 
  in the polarization plane  
  for the molecular model potential shown in Fig.~\ref{fig3}(a) and the dimer model potential shown in Fig.~\ref{fig3}(d), reproduced for convenience in the inset. The data in (a) were used to plot the grey distribution in Fig.~\ref{fig3}(b), while the data in (b) were used for the green distribution in Fig.~\ref{fig3}(b).
  (c, d) Example trajectory (blue line) of an electron tunneling from the up-field site of the model dimer in position space (c) and momentum space (d). The green dot marks the beginning of the trajectory while the red dot marks the final momentum value,  indicated both in (b) and (d). 
  The color density in (c) depicts the Coulomb potential of the doubly charged dimer. The dashed arrow in (b) denotes the clockwise  rotation of the laser electric field vector.    }
  \label{fig4}
\end{figure}

The simulation for the molecular-type potential with the shorter internuclear distance  resulted in a symmetric MFPAD with its symmetry axis around $\alpha=85^\circ$ [see gray line in Fig.\,\ref{fig3}(b)], similar to the measured MFPAD in Fig.~\ref{fig2}(a). 
This case also applies to the dimer-case when the  potoelectrons are emitted from the down-field potential well, as in both cases the electrons tunnel directly into the continuum [cf. Fig.~\ref{fig3}(a)]. 
In contrast, the simulated  angular distribution of photoelectrons emitted from the up-field potential well of the dimer-like potential with the larger intermolecular separation features a stronger rotation and a clearly visible distortion, in agreement with the measurement, cf. Fig.\,\ref{fig3}(b).  The angular distribution in Fig.\,\ref{fig3}(b) is separated into two cases:  The red line shows the DFPAD for electrons emitted to the left, the green line for electrons emitted to right [cf. the sketches  Figs.~\ref{fig3}(c) and (d)]. 
In both cases the DFPAD in Fig.\,\ref{fig3}(b) shows a strong rotation. 
The simulated photoelectron momentum distributions from which the MFPAD and the green-colored half of the DFPAD in Fig.~\ref{fig3}(b) are calculated are shown in Figs.\,\ref{fig4}(a) and \ref{fig4}(b), respectively.

The simulations, thus, predict a stronger rotation and distortion of the angular distributions of photoelectrons emitted from the upper potential [red and green lines in Fig.~\ref{fig3}(b)] as compared to photoelectrons emitted from the lower potential well [gray line in Fig.~\ref{fig3}(b)]. 
By a detailed analysis of the electrons' trajectories we found that the dominant reason for this stronger rotation is that the photoelectrons emitted from the up-field potential well,   scatter off the neighbouring potential well as they are driven away by the strong laser field.
To visualize this process we show in Figs.\,\ref{fig4}(c) and \ref{fig4}(d) the temporal evolution of an example trajectory in real space and in momentum space, respectively. The green dot depicts the trajectory's starting conditions (space and momentum). The red dot indicates the final momentum [also indicated in Fig.~\ref{fig4}(b)]. For trajectories that make up the red half of the simulated DFPAD in Fig.~\ref{fig3}(b) the situation is reversed and the trajectories emitted from the right well scatter on the left potential well.
Thus, based on our semiclassical trajectory simulations we attribute the rotation and distortion of the measured DFPADs in Figs.~\ref{fig2}(b) and (c) to scattering of photoelectrons emitted from the up-field potential well on the charge in the opposing potential well of the partner molecule in the dimer.

\section{Information retrieval from the DFPADs}

We now turn to discussing the information that is contained in DFPADs measured with strong, elliptically polarized laser fields and how it can be extracted. 
It was shown above that a DFPAD can be quite different from the corresponding MFPAD. Specifically we have shown that the nearby presence of a molecular ion results in a notable rotation and deformation of the photoelectron angular distribution from the dimer. 
%
In the following we will show for the example of the heterodimer N$_2$-O$_2$ that, despite the distortions that are introduced by the presence of the  neighbouring molecular ion in the DFPAD, it is still possible to extract from it detailed information about the ionization process in the elliptically polarized pulse and the nature of the distortion due to the neighouring ion. 
%


To start the discussion, we note that double ionization and fragmentation of the N$_2$-O$_2$ dimer in the pump-probe scheme of our experiment can proceed via two pathways. In the first pathway, the  N$_2$ molecule is ionized during the linearly polarized pump pulse and subsequently the O$_2$ molecule becomes ionized by the elliptically polarized probe pulse, see Equ.~(\ref{pthw1}). During the second pathway the sequence of ionization is reversed, see Equ.~(\ref{pthw2}).
\begin{eqnarray}
\rm{N_2\mbox{-}O_2} &\rightarrow \rm{N_2^+\mbox{-}O_2} &\rightarrow \rm{N_2^+\mbox{-}O_2^+} \label{pthw1}\\
\rm{N_2\mbox{-}O_2} &\rightarrow \rm{N_2\mbox{-}O_2^+} &\rightarrow \rm{N_2^+\mbox{-}O_2^+} \label{pthw2}
\end{eqnarray}
%
In both pathways the ionization process during the pump pulse takes place within a neutral dimer. 
The ionization dynamics of the second molecule during the probe pulse, in contrast, takes place in the vicinity of a molecular ion. The presence of this ion may modify the ionization process of the second molecule. 
Which ionization step will be modified more strongly by the presence of the neighbouring ion as compared to the neutral case, the one of O$_2$ or of N$_2$?


To explore this question, one should evaluate the probability ratio of pathways~(\ref{pthw1}) and (\ref{pthw2}).
Is it possible to extract this ratio from the angular distribution of the second electron that is emitted from the heteromolecular dimer? 
As we will show in the following, the answer to this question is yes. In our experiment we measure the electrons and ions emitted in coincidence. Therefore, we can display the PAD in the dimer frame, which results in a DFPAD. The crucial point is, however, that for a heterodimer, the DFPAD can be further refined by not only fixing the dimer-axis along the, e.g., horizontal, axis, but also by \textit{orienting} the dimer with respect to the molecular species. While in Fig.\,\ref{fig2}(c) the orientations (N$_2$ ejected to the left/right or O$_2$ ejected to the left/right) were not considered and therefore Fig.\,\ref{fig2}(c) is a superposition of both cases, the DFPAD displayed in Fig.~\ref{fig3}(e) is plotted such that N$_2$ is always ejected to the right. 


Under this condition the oriented DFPAD can be separated into two halves by resorting to the concept of angular streaking \cite{Maharjan2005, Hanus2019, Hanus2020}. 
For the clockwise helicity of the elliptical laser field used in our experiment electrons are streaked into the upper half [colored in red in Fig.~\ref{fig3}(e)] when the laser electric field vector points from  O$_2$  to N$_2$, cf. sketch in Fig.~\ref{fig3}(c).
In contrast, the green-colored lower half in Fig.~\ref{fig3}(e) corresponds to electrons that are emitted when the laser field vector points from N$_2$ to O$_2$, cf. sketch in Fig.~\ref{fig3}(d).
If we compare the shapes of the two halves in the oriented DFPAD in Fig.\,\ref{fig3}(e)
we notice that they are distinctly different. To highlight this difference we overlaid in Fig.\,\ref{fig3}(e) the non-oriented DFPAD from Fig.~\ref{fig2}(c) by a red line. 
By comparison of the oriented DFPAD with this red line it becomes obvious that the upper, red-colored half comprises of significantly more electrons. Thus, we can conclude that obviously the second ionization step happens with a higher probability when the laser electric field vector points from O$_2$ site to N$_2$ [sketch in Fig.~\ref{fig3}(c)]. 
This does, however, not imply that the second electron is more likely emitted from O$_2$ in the down-field well of the dimer. 
Thus, to quantify the probability ratio and to determine how many electrons are emitted from O$_2$ respectively N$_2$, it is not simply possible to integrate the upper and lower halves of the DFPAD in Fig.~\ref{fig3}(e).

\new[R2-4 ]{The reason is that for dimers the laser field driven trajectories of the emitted electrons are strongly influenced by the potential of the neighbouring molecular ion, as we have discussed above. 
Consquently, electrons that are emitted when the electric field vector points into a certain direction may not be detected under an angle of 90$^\circ$ to the field direction, as it is predicted by angular streaking based on the  strong-field approximation (SFA) \cite{Faisal1973, Reiss1980} that neglects the influence of any ionic potential on the final momentum of the photoelectron. }
But due to scattering on the neighbouring molecular ion, in particular electrons from the up-field potential well may be streaked into a distinctively different direction. This leads to deformations, but mainly to a global rotation of the overall DFPAD as compared to the SFA prediction. This was clearly shown by the results of our simulations, cf. Fig.~\ref{fig3}(b). 
Based on the finding described above that the dominant deviation from the SFA prediction is a rotation of the angular distribution of photoelectrons emitted from the up-field potential well along the clockwise rotation of the laser electric field vector, we can, however, separate the scattered from the non-scattered electrons. 

To understand the rotation due to scattering, we turn to the simulated results in Fig.~\ref{fig3}(b). This figure shows that electrons that are emitted 
from the down-field potential well  directly into the continuum (gray line), will only be marginally affected by the Coulomb binding  potential. The overall effect is that they, when emitted dominantly around  the peaks of the laser field along the main axis of the laser polarization ellipse (at $0^\circ$  and $180^\circ$), will be deflected only by a few degrees 
further along the laser field vector rotation direction (clockwise in our case) than the 90$^\circ$ predicted by the SFA. As a consequence, their PAD shows peaks around $85^\circ$  and $265^\circ$.
In contrast, electrons that are emitted from the up-field potential well are scattered on the neighbouring well, leading to a much larger rotation of the angular distribution. 
Our model, which reproduced the measured DFPADs reasonably well, predicts that the additional rotation due to scattering is about 40$^\circ$, such that the angular distribution of these electrons shows peaks around $50^\circ$  and $230^\circ$, see red and green lines in Fig.~\ref{fig3}(b).
Based on these values we divide each half of the DFPAD in Fig.~\ref{fig3}(e) into two segments using 60$^\circ$ respectively 240$^\circ$ as the borders. This results in the four segments denoted I to IV, indicated by darker and lighter red and green shading in Fig.~\ref{fig3}(e). 
\new[R2-1b ]{We note that the scattered and non-scattered regions in the DFPAD certainly overlap to some extent. It is therefore intrinsically impossible to unequivocally disentangle these regions using sharp angular limits. As a result, precise quantitative results cannot be expected from such a division of the DFPAD. The purpose of the analysis described below is, however, not to provide quantitative results, but to demonstrate a possible approach to disentangling different processes in measured DFPADs.
We think that this approach could be extended to a qualitative level using a more advanced division of the DFPAD.

In the following we discuss how the regions I to IV defined by the sharp angular limits can be used to estimate the emission percentage of the second electron from the up-field and down-field potential wells. And subsequently, how from this information an estimate for the branching ratio between pathways (\ref{pthw1}) and (\ref{pthw2}) can be obtained.
}
Region I in the upper half of the DFPAD corresponds to electrons that were scattered after their release from the up-field potential well in the situation depicted in Fig.~\ref{fig3}(c). Thus, they can be attributed to emission from N$_2$. Electrons in region II were not scattered. Therefore, they were emitted from the down-field well, i.e., from O$_2$. With the same logic, region III contains scattered electrons emitted from O$_2$, while region IV corresponds to non-scattered electrons emitted from N$_2$.
With this assignment, it is now simple to estimate the percentage of the two pathways described  by  Equs.~(\ref{pthw1}) and (\ref{pthw2}): Regions I and IV contain electrons emitted from N$_2$ during the probe pulse (second ionization step), thus, represent pathway (\ref{pthw2}). Regions II and III contain electrons emitted from O$_2$ during the second ionization step and, thus, represent pathway (\ref{pthw1}). By integrating the numbers in these regions we obtain that 54\% of the total counts are due to pathway (\ref{pthw1}), while only 46\% of all events follow pathway (\ref{pthw2}).
\new[R2-1b ]{We emphasize once more that these numbers are of no specific value. It is the demonstration of an approach that we are interested in rather than obtaining an exact result. 
We have, however, checked by small variations of the angular limits in the DFPAD that the numbers are relatively robust with respect to the specific choice of the limits. }
Hence, despite the intrinsic uncertainties we would like to conclude that 
in our experiment the pathway when the O$_2$ is ionized in the second ionization step is somewhat more likely. This can be interpreted such that the two ionization steps are not completely independent of each other. 

Which factors could be responsible for this ionization behaviour?
An ionization process that favors electron emission from a specific potential well in a molecule exposed to a strong field is enhanced ionization (EI) \cite{Codling1989, Xie2014a}. It has been shown that this molecular ionization process takes place in almost the same manner also in dimers \cite{Wu2012b}. However, EI favors electron emission from the up-field well, while in our experiment the more probable electron emission from O$_2$ takes place for the down-field well. 
Thus, EI cannot be hold responsible for the ionization dynamics observed in our experiment.
A more likely reason for the observed preponderance of pathway (\ref{pthw1}) is 
the influence of the structure of the heteromolecular dimer on the ionization probability during the two-step double-ionization in our double-pulse experiment. 
The ionization probability of an isolated molecule with a given valence electron configuration and binding energy depends mainly on the orientation of the molecule with respect to the laser field. 
If two molecules are bound together by van der Waals forces, a number of different dimer configurations become possible \cite{Hoshina2012, Gong2013}, for example T-shaped, X-shaped, parallel,  linear, etc. 
Thus, 
not only the valence structure of the two molecules and their different ionization potentials, but also the dimer geometry determines which dimers from the randomly oriented ensemble in the utrasonic gas jet are preponderantly ionized. 
Therefore, in a double pulse scheme, particularly when the two pulses have different polarization states, as in our experiment, the probability of a particular pathway is determined by many convoluted parameters. 
A detailed analysis of the 
hypothesis that the dimer structure is responsible for the slightly favorable ionization of O$_2$ during the second ionization step is, however, beyond the present work and must be left for future work.




\section{Summary}

In summary, we described the results of experiments and simulations performed with the aim of exploring whether and how information extraction from laser-generated photoelectron angular distributions (PADs), a standard method in atomic and molecular physics, can be extended to the case of molecular compounds. To this end we have studied strong-field double-ionization of homo- and heteromolecular dimers of O$_2$ and N$_2$ formed by van der Waals binding forces. To distinguish the two ionization steps we applied two delayed ultrashort intense laser pulses, where the first pulse was linearly polarized and the second one elliptically polarized. In combination with four-body coincidence imaging using a reaction microscope this allowed us to measure dimer-frame photoelectron angular distributions (DFPADs). By a detailed analysis of the DFPAD for the heteromolecular O$_2$-N$_2$ dimer we showed that the PAD of a molecular dimer is deformed and rotated as compared to the PAD of an isolated molecule. With the help of simulations we showed that these distortions are mainly due to scattering of electrons emitted from the up-field potential well 
on the potential of the molecular ion 
formed during the first ionization step. Building on this finding, we demonstrated that by dividing the DFPAD into regions that mainly contain scattered electrons, and regions containing mainly non-scattered electrons, it becomes possible to overcome the complications due to the DFPAD-distortion and to extract information about the ionization pathway.

\new[R2-Bullet4 ]{
The results of our study point to a promising possibility of extracting information about bound electron dynamics induced by a pump pulse from a DFPAD. Such extraction has been the subject of many works on isolated molecules \cite{Lochbrunner2001, Gessner2006, Akagi2009, Staudte2009, Odenweller2011, Spanner2012, Wu2012, Hanus2020}. 
The underlying principle of these efforts is that the angular ionization probability reflects the intra-molecular bound electron density. Thus, the bound electronic dynamics becomes encoded in the measured MFPAD. 
In turn, by a time-resolved measurement of the MFPAD it becomes possible to obtain insight into the electronic dynamics in molecules.

Our work investigates the case of molecular dimers and shows that the resulting DFPAD contains valuable information about the electronic dynamics during laser-dimer interaction. Moreover, we demonstrate a possible route to extract this information.
Thus, if the DFPADs are measured with a variable delay between the first and second ionization step (rather than only for one value of the delay as in our work), the femtosecond evolution of dynamics in dimer-molecules induced by the pump pulse can be extracted. 
}





\acknowledgments
This work was supported by the Austrian Science Fund (FWF), Grants No. 
P28475-N27  
and P30465-N27.   



%

\end{document}